\documentclass[a4paper]{article}
\usepackage{graphicx,subfigure,amssymb,amsfonts}
\newcommand{\req}[1]{(\ref{#1})}
\begin{document}

\title{Algorithms of maximum likelihood data clustering with
applications} 

\author{Lorenzo Giada$^*$ and Matteo Marsili$^+$ \\
{\small$^*$ \em Max Planck Institut f\"ur Kolloid- und
Grenzfl\"achenforschung, 14424 Potsdam, Germany} \\
{\small$^+$\em Istituto Nazionale per la
Fisica della Materia (INFM), Trieste-SISSA Unit,}\\ {\small \it V. Beirut 2-4,
Trieste I-34014}} 

\date{\today} 
\maketitle
\begin{abstract}  
We address the problem of data clustering by introducing an
unsupervised, parameter free approach based on maximum likelihood
principle. Starting from the observation that data sets belonging to
the same cluster share a common information, we construct an
expression for the likelihood of any possible cluster structure. The
likelihood in turn depends only on the Pearson's coefficient of the
data. We discuss clustering algorithms that provide a fast and
reliable approximation to maximum likelihood configurations. Compared
to standard clustering methods, our approach has the advantages that
{\em i)} it is parameter free, {\em ii)} the number of clusters need
not be fixed in advance and {\em iii)} the interpretation of the
results is transparent. In order to test our approach and compare it
with standard clustering algorithms, we analyze two very different
data sets: Time series of financial market returns and gene expression
data. We find that different maximization algorithms produce similar
cluster structures whereas the outcome of standard algorithms has a
much wider variability.
\end{abstract}

\vspace{2cm}
Keyword
Dataclustering; Econophysics; Gene expression

\section{Introduction}
One aspect of the Information Technology revolution is that huge
amounts of data have become available. For example, every transaction
in many financial markets is recorded and bio-informatics technology
allows us today to monitor genome wide gene expression. These
informations may allow for a quite detailed description of such
complex systems as a financial market or a cell, to quote just two
examples. However our understanding of a complex systems is, in many
cases, limited by our ability to efficiently organize massive streams
of information.  As a result, methods for handling, organizing or
mining large data sets have become of great practical importance.

Data clustering deals with the problem of classifying a set of $N$
objects into groups so that objects within the same group are more
similar than objects belonging to different groups. Each object
is identified by a number $D$ of measurable features: Hence
object $i=1,\ldots, N$ can be represented as a point $\vec
x_i=(x_i^{(1)},\ldots,x_i^{(D)})$ in a $D$ dimensional space. Data
clustering aims at identifying clusters as more densely populated
regions in this vector space. More precisely, a configuration of
clusters is represented by a set $\mathcal{S}=\{s_1,\ldots,s_N\}$
of integer labels, where $s_i$ is the cluster to which object $i$
belongs.

We focus on unsupervised approaches, which do not require any further
information apart from that contained in the data. These are of great
practical importance specially in the presence of huge data sets. We
assume, in other words, that the data set is homogeneous: all features
are equally important or are already appropriately weighted. 

The classic approaches to data clustering are partitioning methods and
hierarchical clustering.  Partitioning methods are based on two
elements: {\em 1)} a distance between objects, which allows one to
measure their similarity and {\em 2)} a cost function whose minima
correspond to ``optimal'' clustering configurations. For example, a
typical K-means (KM) approach takes as cost function the sum of
squared distances of objects to the centroid $\vec X_s$ of the cluster
in which they are classified \cite{booko}:
\begin{equation}
H_{KM}\{\mathcal{S}\}=\sum_s\sum_{i: s_i=s}\|\vec x_i-\vec X_s\|^2,~~~\vec
X_s=\frac{1}{n_s}\sum_{i: s_i=s}\vec x_i
\label{HKM}
\end{equation}
where $n_s=\sum_{i: s_i=s} 1$ is the number of objects in cluster $s$,
and $\| \vec x \|^2=\sum_j (x^{(j)})^2$ defines the Euclidean
distance.

Apart from cases where a cost function is naturally suggested by the
problem itself\footnote{The typical example is that of deciding the
location of $K$ distribution centers which should serve $N$ cities. It
is natural to look for the locations which minimize the sum of
distances.}, the choice of the cost function or of a distance is an
element of arbitrariness. A further problem of these approaches is
that one needs to predefine the number $K$ of clusters from the
beginning. Note, for example that $H_{KM}$ of Eq. (\ref{HKM}) attains
its minimal value $H_{KM}=0$ when each object is in a different
cluster ($n_s=1$). The number $K$ of clusters should then be fixed
{\em a priori} or one has to introduce a $K$-dependent term in
the cost function. This is a further element of arbitrariness.

The problem of finding the cluster structure which minimizes the cost
function may be quite difficult: depending on the form
of the cost function and on the data set the "cost landscape" can be
either simple, with a single minimum which is easily accessible
dynamically, or complex, with many metastable minima. Data clustering
methods also differ for the specific algorithms used to reach a
(local) minimum. 

A second, very popular approach to data clustering -- called
hierarchical clustering -- is based on the definition of a distance
between objects and clusters of objects and a very simple algorithm:
Given a configuration with $K>1$ clusters, it merges the two
closest clusters into a single one. In this way, starting from the
configuration with $K=N$ clusters, the algorithm generates a sequence
of configurations as $K$ varies from $N$ to $1$. This sequence of
configurations and their hierarchic organization, can be represented
by a convenient and compact graphical tool called dendrogram
\cite{booko}. The minimal spanning tree algorithm -- also called
Single Linkage (SL) -- the Average Linkage (AL) and the Centroid
Linkage (CL) algorithms are examples of hierarchical clustering
methods\footnote{These algorithms differ in the way the distance
between clusters is computed: In SL (AL) the distance between clusters
$s$ and $r$ is the minimal (average) distance between items in cluster
$s$ and items in cluster $r$; in CL the distance $\|\vec X_s-\vec
X_r\|$ between centroids is used.}.  Rather than a single cluster
structure, this algorithm provides a hierarchic sequence of cluster
structures. The choice of the best cluster structure is left
arbitrary. Applications of this approach to real world data is
discussed in Refs. \cite{Eisen,Mantegna}.

Many other approaches to data clustering have been proposed: For
example Refs. \cite{SVDbio,SVDfin} used singular value decomposition
to identify clusters. Identifying principal components with cluster
structures imposes an orthogonality constraint between clusters which
may be unnatural. Expectation Minimization is a further approach
\cite{EM} where the density of points is modeled as a mixture of
Gaussians whose centering and scale parameters are fit by maximum
likelihood. 
Kohonen et al. \cite{Tamayo} have proposed an
algorithm based on Self-Organizing Maps (SOM) whereas Blatt {\em et
al.} proposed the Super-Paramagnetic Clustering (SPC) method
\cite{SPC}. The latter is based on a mapping to an interacting
particle system whose ``magnetic" properties describe the cluster
structure of data and has been applied to a range of problems (see
e.g. \cite{SPCbio,GLD,SPCfin}). These methods rely on {\em ad hoc}
definitions of a dynamics (SOM) or of the particle-particle
interaction (SPC) which are tuned by several parameters or functions.

In the following sections we briefly review yet a different approach
to dataclustering that we recently proposed \cite{GM}, we describe
possible algorithms based on it, and we finally compare the outcomes
with those of standard data clustering methods, as those described
above.

\section{Maximum likelihood data clustering}

We have devised \cite{GM} a fully unsupervised, parameter free
approach to data clustering which derives from a maximum likelihood
(ML) principle.  The key idea is that objects are similar if they have
something in common.  In a correct classification, objects belonging
to the same cluster should share a common component:

\begin{equation}
\vec x_i=
g_{s_i}\vec\eta_{s_i}+\sqrt{1-g_{s_i}^2}\vec\epsilon_i.
\label{ansatz}
\end{equation}

\noindent
Here $\vec x_i$ is the vector of features of object $i$, normalized so
that $\sum_t x_i^{(t)}=0$ and $||\vec x_i||^2=\sum_t [x_i^{(t)}]^2=D$
for all $i=1,\ldots,N$\footnote{Equivalently, one may generalize
Eq. \req{ansatz} by adding a constant term $r_i\vec 1$ and a scale
factor $\sigma_i$. The maximum likelihood estimates of these
parameters are the mean and the variance. Subtracting the mean and
rescaling by the variance, leaves us with the normalized data set.
The parameters $r_i$ and $\sigma_i$ are irrelevant as far as the
cluster structure is concerned. The latter indeed only depends on the
``internal structure'' of correlations.}; $s_i$ is the label of the
cluster to which it belongs. $\vec \eta_s$ is the vector of features
of cluster $s$ and $g_s$ tunes the similarity of objects within
cluster $s$: For $g_s=1$ all objects with $s_i=s$ are identical
whereas when $g_s$ is small objects are very different. The cluster
index $s$ ranges from $1$ to $N$ in order to allow also for the case
of $N$ clusters of one object each. $\vec\epsilon_i$ describes the
deviation of the features of object $i$ from the cluster's features
and measurement errors. We take Eq. \req{ansatz} as a statistical
hypothesis and assume that both $\vec \eta_s$ and $\vec \epsilon_i$
(for all $i,s=1,\ldots,N$) are Gaussian vectors with zero mean and
variance $E[(\eta_s^{(t)})^2]=E[(\epsilon_i^{(t)})^2]=1$. For any
given set of parameters $(\mathcal{G,~S})=(\{g_s\},\{s_i\})$ it is
possible \cite{GM} to compute the probability (density) $P(\{\vec
x_i\}|\mathcal{G,~S})$ of observing the data set $\{\vec x_i\}$ as a
realization of Eq. \req{ansatz}. From this it is possible \cite{GM} to
compute the likelihood\footnote{Note that an assumption on the {\em
prior probability} $P(\mathcal{G,~S})$ is invoked by Bayes formula in
this passage. We take $P(\mathcal{G,~S})=\hbox{const}$ which means
that every cluster structure $\mathcal{S}=\{s_i\}$ is {\em a priori}
equiprobable.}  $P(\mathcal{G,~S}|\{\vec x_i\})$. It turns out
\cite{GM} that the resulting expression depends only on the Pearson's
coefficient

\begin{equation}
C_{i,j}=\frac{\vec x_i\cdot\vec x_j}{\sqrt{||\vec x_i||^2||\vec x_j||^2}}.
\end{equation}
For a given cluster structure $\mathcal{S}$, the likelihood is maximal
when the parameters $g_s$ take the values

\begin{equation}
g_s^\star=\sqrt{\frac{c_s-n_s}{n_s^2-n_s}}
\end{equation}

\noindent
if $n_s>1$ and $g_s^\star=0$ if $n_s\le 1$. Here $n_s$ is the number
of objects in cluster $s$ and

\begin{equation}
c_s =\sum_{i=1}^N\sum_{j=1}^NC_{i,j}\delta_{s_i,s}\delta_{s_j,s}.
\label{nscs}
\end{equation}
Note indeed that when $c_s\approx n_s$, as for uncorrelated objects,
then $g_s^\star\approx 0$ whereas if objects are very similar, $c_s\approx
n_s^2$ and $g_s^\star\approx 1$. 

The maximum likelihood of structure $\mathcal{S}$ can be written as
$P(\mathcal{G^\star,~S}|\{\vec x_i\})\propto e^{D\mathcal{L}_c(\mathcal{S})}$,
where the log-likelihood per feature $\mathcal{L}_c$ is given by

\begin{equation}
\mathcal{L}_c(\mathcal{S})=\frac{1}{2}\sum_{s:~n_s>1}
\left[\log
\frac{n_s}{c_s}+(n_s-1)\log\frac{n_s^2-n_s}
{n_s^2-c_s}\right].
\label{Hc}
\end{equation}
This depends on the original data through the coefficients $c_s$ of
Eq. \req{nscs}. The function $\mathcal{L}_c$ provides a likelihood
measure for cluster structures. The ML structure is
that which maximizes $\mathcal{L}_c$.

There are several interesting features of $\mathcal{L}_c$:
\begin{itemize} 
\item If the objects are unrelated ($g_s^\star=0$ or $c_s=n_s$) or if they
are classified in singleton clusters ($n_s=1$ for all $s$) we find
$\mathcal{L}_c=0$. Loosely speaking, $\max_\mathcal{S}\mathcal{L}_c(\mathcal{S})$
measures the amount of structure present in the data-set.
\item The maxima of $\mathcal{L}_c$ do not necessarily coincide with a
single cluster containing all objects -- as in the SPC approach of
Ref. \cite{SPC} -- nor with the configuration with all objects in
different clusters -- as for $- H_{KM}$\footnote{Note that
Eq. \req{HKM} can be written as $H_{KM}=2\sum_{s:n_s>0}(n_s-c_s/n_s)$.}
\item $\mathcal{L}_c$ does not depend on any parameter.
\item The number $K$ of clusters is not fixed {\em a priori}. Rather
it is predicted.
\item The interpretation of the results is transparent in terms of the
model \req{ansatz}. 
\end{itemize}

Eq. \req{ansatz} may not be the most appropriate description for a
particular data set. In much the same way, a straight line may not be
the best description of a set of points on a plane. Still in this
example, least squares provide an unambiguous method to compute the
coefficients and a statistical measure of the goodness of fit. The
same is true for our method: A sharp maximum of $\mathcal{L}_c/N$
indicates a robust cluster structure whereas when $\mathcal{L}_c/N$ is
small and has several local maxima the cluster structure is not
statistically significant. 

The statistical hypothesis Eq. \req{ansatz} is specially helpful for
high dimensional data sets ($D\gg 1$) where geometric intuition
becomes problematic. From the point of view of the computational cost,
the dimensionality $D$ of the data set enters only in the calculation
of the matrix $C_{i,j}$. For small $D$ methods based on geometry and
visual inspection may be preferrable. 

Our approach has some similarity with Expectation Maximization data
clustering \cite{EM}, which is also based on likelihood
maximization. However the statistical hypothesis in Ref. \cite{EM} is
very different from Eq. \req{ansatz}.

\section{Data clustering algorithms}

The function $\mathcal{L}_c(\mathcal{S})$ is a measure of likelihood for
cluster structures and it can be used to compare cluster structures
produced by standard methods. For example, it allows one to compare
classifications with a different number of clusters and hence to
select the optimal number of clusters in KM, SL, AL or CL
algorithms. We shall compare later the cluster structure found in this
way with those found with algorithms based on $\mathcal{L}_c$ itself.
Most importantly in fact, $\mathcal{L}_c$ can be used as the basis for
clustering algorithms.

A very powerful method to find ML configurations is simulated
annealing (SA) \cite{SA} with cost function $-\mathcal{L}_c$. This is
simply implemented via Metropolis dynamics \cite{Metropolis} on the
dynamical variables $s_i$ with progressively decreasing (fictitious)
temperature $T$. This method produces a cluster configuration, which,
if the annealing schedule is appropriately chosen, has a good chance
of being that of maximum likelihood.  The behavior of the system as a
function of the fictitious temperature is similar to that described in
Ref. \cite{SPC}. However, the SPC method (in its original formulation
\cite{SPC}) yields a trivial configuration with a single cluster in
the limit $T\to 0$. This is a consequence of the mapping of the data
set into a particle system with ferromagnetic interactions,
corresponding to positive correlations. This forces one to study
intermediate temperatures\footnote{Modifications to take into account
negative correlations are suggested in Ref. \cite{SPCfin}.}.  The ML
configuration found in our method when $T\to 0$ is already
non-trivial.  Further insight on the cluster structure can be obtained
by studying the system at finite $T$, as shown in Ref. \cite{GM}. For
simplicity however, we will limit the present discussion to the $T\to
0$ limit.

A second possibility is that of using a deterministic maximization
(DM) technique. We discuss below a very simple algorithm: Given a
configuration $S$, for all $i=1,\ldots,N$ (in some fixed order)
propose all moves $s_i\to s$ and compute the corresponding variation
$\Delta_{s_i\to s}$ of $\mathcal{L}_c$; Find $s_i'={\rm arg}\max_s
\Delta_{s_i\to s}$ and set $s_i=s_i'$. Repeat these steps until no
change occurs. This will lead in the end to a local maximum of
$\mathcal{L}_c$ which we shall call the $\mathcal{S}_{DM}$
configuration.

Finally we propose a merging algorithm (MR) in the spirit of
hierarchical clustering based on $\mathcal{L}_c$.
\begin{enumerate}
\item[0)] Start from $N$ clusters composed of one object each
(e.g. $s_i=i$ $\forall i=1,\ldots,N$).
\item[1)] At each step of the algorithm merge two clusters
into a single one in such a way that the cost $\mathcal{L}_c$ of the
resulting configuration is minimal. 
\item[2)] Repeat step 1) $N-1$ times until the
configuration with a single cluster is reached. 
\end{enumerate}
Let $\mathcal{S}_{MR}$ be the state with maximal $\mathcal{L}_c$ found with
this algorithm. $\mathcal{S}_{MR}$ is not necessarily a local maximum of
$\mathcal{L}_c$ with respect to single ``spin-flip'' moves $s_i\to s_i'$,
while $\mathcal{S}_{SA}$ and $\mathcal{S}_{DM}$ are local maxima. Hence, we
expect that $\mathcal{L}_c(\mathcal{S}_{MR})\le\mathcal{L}_c(\mathcal{S}_{DM})\le
\mathcal{L}_c(\mathcal{S}_{SA})$. On the other hand, the hierarchical
clustering algorithm is definitely faster than both of the others:
Even if it delivers only an approximation to the maximum likelihood,
faster execution can be crucial when dealing with huge data sets.

In addition, the hierarchical clustering technique MR offers a very
convenient graphical representation of the cluster structure in terms
of dendrograms. Usually dendrograms report the tree structure as a
function of a resolution length. As this length increases, clusters
are merged one by one until a single cluster remains when the
resolution exceeds the maximal distance $\max_{i,j}\|\vec x_i-\vec
x_j\|$ between data points. We draw the dendrogram as a function of
the log-likelihood per feature $\mathcal{L}_c$. At each merging step $s,
r\to q$ let $\ell_s$, $\ell_r$ be the contribution of clusters $s$ and
$r$ respectively to $\mathcal{L}_c$ before merging and $\ell_q$ be the
corresponding cluster log-likelihood of the merged cluster $q$.  In
the dendrogram, we merge two points at height $\ell_s$ and $\ell_r$
into a single point at height $\ell_q$. There are three types of
branching points: {\em i)} if $\ell_q>\ell_s+\ell_r$ the step
increases the likelihood of the configuration {\em ii)} if
$\ell_q<\ell_s+\ell_r$ but $\ell_q>\max(\ell_s,\ell_r)$ the global
likelihood decreases but the cluster likelihood increases. Finally
{\em iii)} steps with $\ell_q<\max(\ell_s,\ell_r)$ represent unlikely
merging moves.

\section{Applications and discussion}

We discuss the application of the above ideas to two different data
sets: The first consists of daily returns of the NYSE from January the
1st 1987 to March the 30th 1999. We consider $N=1000$ of the assets
traded most frequently throughout this period \cite{supplement}. If
$p_i(t)$ is the opening price of asset $i$ in day $t$, we set
\begin{equation}
x_i^{(t)}=\frac{\log p_i(t)/p_i(t-1) - r_i}{\sigma_i}
\label{norm_returns}
\end{equation}
where $r_i$ and $\sigma_i$ enforce normalization ($\vec x_i\cdot\vec 1=0$
and $\|\vec x_i\|^2=D$). Actually it is convenient to introduce a
further linear transformation in order to eliminate the so called
``market mode'': $\vec x_i'=\alpha_i(\vec x_i-\vec x_{\rm avg})$ where
$\vec x_{\rm avg}=\sum_i \vec x_i/N$ so that $\sum_i \vec x_i'=0$
($\alpha_i$ is a coefficient to restore $\|\vec x_i'\|^2=1$). 

Secondly we discuss the gene expression data set measured in
\cite{Spel} and already studied in Refs. \cite{Eisen,SVDbio,SPCbio}
with a variety of techniques. The data measure the expression of the
2467 genes of known function of the yeast {\em Saccharomyces
cerevisi\ae}, whose complete genome has been sequenced. In particular,
we focused on the 18 time values coming from the alpha-factor
block-and-release experiment in \cite{Spel}; these correspond to
approximately two cell cycles. {Also in this case, we take the
logarithm of the value of the gene expression, so that the data
distribution is closer to a Gaussian, and normalize the sets as
required by Eq. \ref{ansatz}.}

In both cases the data set is homogeneous by definition: features
characterize the system -- the market or the cell -- at different
times and there is no {\em a priori} reason to weight features
differently\footnote{Note however that some major crashes such as
black Monday of October 1987 are included in the financial data
set.}. The main objective of data clustering for these data sets is
that of identifying groups of assets with a similar market behavior or
groups of genes with a similar function.  This information is of
considerable interest for risk management strategies and for the
understanding of biological functions respectively in the two
cases. We shall not enter here into details of either finance or
genetics, but rather
focus on the properties of data clustering algorithms for these two
case studies. {On one hand we are interested in identifying fast
clustering algorithms, on the other it is desirable that the resulting
cluster structures does not depend strongly on the detailed algorithm
used. In few words, one would like to identify fast algorithms which
provide results which are consistent with more elaborate and time
consuming ones.}  Note that the two data sets are very different in
nature: For market data time series are very long ($D=3114$) and
correlations are very weak while gene expression is recorded for few
($D=18$) time points but correlations are very strong. Thus they
constitute a good testing ground for our analysis.

The cluster structure which emerges from the minimization of $\mathcal{
L}_c$ are quite meaningful: For market data we find a strong overlap
with the classification of stocks in sectors of economic activity. In
the gene expression data set, as in Ref. \cite{SPCbio}, our method
identifies clusters related to biological functions.  This information
is preserved more or less in the ML cluster structures found with
different methods.

\begin{figure}[t]
\centerline{\includegraphics[width=6cm,height=9cm,angle=270,clip]{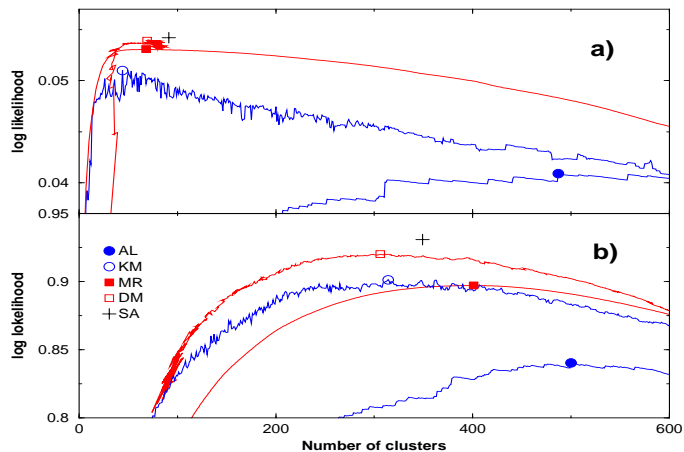}}
\caption{Comparing methods: The log-likelihood $\mathcal{L}_c$ is plotted
against the number of clusters for the different methods discussed in
the text. Points corresponds to the maximal likelihood
configurations. The upper plot (a) refers to financial data, the
bottom one (b) refers to gene expression data. Configurations KM and
DM are obtained by deterministic minimization of $H_{KM}$ and $-\mathcal{
L}_c$ respectively, starting from all configurations generated by AL
and MR respectively. $K$ is held fixed in the minimization of $H_{KM}$
but not in the maximization of $\mathcal{L}_c$. Hence DM can have a
number of clusters different from that of the starting MR
configuration.}
\label{figenergy}
\end{figure}

\begin{table}[t]\center
\caption{Number of clusters $K$, and likelihood per data for the
maximum likelihood cluster structures obtained with different
methods for the gene expression and the financial datasets.}
\begin{tabular}{l|cc|cc}
\multicolumn{1}{c}{~~} &
\multicolumn{2}{c}{Gene expression} &
\multicolumn{2}{c}{Financial market}\\
Method                        & $K$ & $\mathcal{L}_c/N$ & $K$ & $\mathcal{L}_c/N$\\
\hline
Single Linkage (SL)		& $1788$& $0.3046$ &$885$ & $0.0285$  \\
Centroid Linkage (CL)		& $615$	& $0.7806$ &$910$ & $0.0235$  \\
Average Linkage (AL)    	& $500$ & $0.8404$ &$487$ & $0.0409$  \\
K-means (KM)            	& $314$ & $0.9014$ & $44$ & $0.0510$  \\
\hline 
$\mathcal{L}_c$ Merging (MR)      	& $401$ & $0.8972$ & $68$ & $0.0531$  \\
Determ. min $\mathcal{L}_c$ (DM)  	& $306$ & $0.9202$ & $69$ & $0.0539$  \\
Sim. Anneal. $\mathcal{L}_c$ (SA) 	& $349$ & $0.9308$ & $91$ & $0.0542$  \\
\hline
\end{tabular}
\label{table1}
\end{table}

\begin{figure}[t]
\centerline{\includegraphics[width=7cm,height=10cm,angle=270,clip]{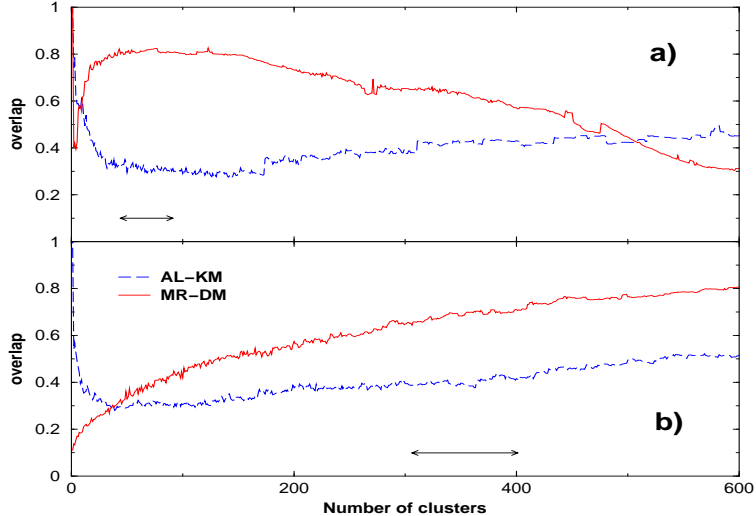}}
\caption{Comparing hierarchical clustering and deterministic
minimization algorithms: From the configuration AL (MR) with $K$
clusters found by hierarchical clustering we run deterministic
minimization of $H_{KM}$ (respectively $-\mathcal{L}_c$) and find the
configuration KM (DM). Overlaps are computed as
$\sqrt{O(X|Y)O(Y|X)}$ (see caption of Table 2) where
(X=AL, Y=KM) or (X=MR, Y=DM). The upper plot (a) refers to financial
data whereas the bottom one (b) refers to gene expression data. Note
that KM has the same number of clusters $K$ as AL but $K$ can be
different for DM and MR because $K$ need not be fixed in the
maximization of $\mathcal{L}_c$. The arrows in the plot indicate the
range of $K$ for ML configurations.}
\label{figovlp}
\end{figure}

Standard hierarchical clustering methods (SL, CL and AL) give ML
configurations which differ considerably from that found with other
methods (see Ref. \cite{supplement} for details). Table
\ref{table1} shows that the configurations $\mathcal{S}_{SL}$, $\mathcal{
S}_{AL}$ and $\mathcal{S}_{CL}$ have many more clusters than the others
and a much lower value of $\mathcal{L}_c$. Fig. \ref{figenergy} compares
the log-likelihood of configurations produced in different algorithms
as a function of the number of clusters. The curve corresponding to AL
deviates markedly from those of other methods; the SL and CL curve
(not shown) deviate even more. Not only ML configurations are very
different because of a larger number of clusters. Even configurations
with a fixed number $K$ of clusters produced by the SL, CL and AL
algorithm differ considerably among themselves and from the
corresponding configurations with $K$ clusters found with other
methods (see Fig. \ref{figovlp}).

The closeness of points MR, DM and SA in Fig. \ref{figenergy} suggests
that methods based on $\mathcal{L}_c$ produce consistent results,
irrespective of the algorithm. The closeness of two cluster
structures $\mathcal{S}$ and $\mathcal{S'}$ can be quantified in terms of
overlaps: Consider the intersection $\mathcal{S}\cap\mathcal{S'}$ of the two
cluster structures -- defined so that objects $i$ and $j$ are in the
same cluster in $\mathcal{S}\cap\mathcal{S'}$ only if they are in the the
same cluster both in $\mathcal{S}$ and in $\mathcal{S'}$. We define a
geometric overlap 
\[
O_g(\mathcal{S}|\mathcal{S'})=\frac{|\mathcal{S}\cap\mathcal{S'}|}{|\mathcal{S'}|}
\]
where $|\mathcal{S}|$ is the number of pairs of objects that are in the
same cluster in configuration $\mathcal{S}$. In words, $O_g(\mathcal{
S}|\mathcal{S'})$ is the fraction of pairs of objects that are in the
same cluster in $\mathcal{S'}$ which also belong to the same cluster in
$\mathcal{S}$.  Equivalently, it is the probability that that two
randomly chosen objects which belong to the same cluster in $\mathcal{
S'}$ are also in the same cluster in $\mathcal{S}$.  Tables \ref{table2}
and \ref{table3} report the geometric overlaps between the ML
configurations and confirm the conclusion that methods based on $\mathcal{
L}_c$ yield consistent results. In fact, the overlaps among
configurations MR, DM and SA are generally larger than those in the
sector SL, AL, KM.  These features manifest strongly in weakly
correlated data sets, such as market data, and to a milder extent when
correlations are stronger, as in the gene expression data set. Still
the distribution of cluster sizes produced by SL and AL algorithms is
much more uneven than that found with other methods.

\begin{table}[t]\center
\caption{Geometric overlaps $O_g(\mathcal{S}|\mathcal{S'})$ between cluster
structures for financial data (with $\mathcal{S},~\mathcal{S'}$ being the ML
cluster structures found with the algorithms SL, AL, KM, MR, DM or SA).}
\begin{tabular}{l|ccc|ccc|}
$\mathcal{S'}\backslash \mathcal{S'}$ & SL    &  AL   &  KM   &  MR   &  DM   &  SA\\
\hline
SL              &$1$    &$0.684$& $0.059$ &$0.016$  &$0.019$ &$0.013$ \\
AL 		&$0.736$&$1$    & $0.220$ & $0.081$ &$0.088$ & $0.081$ \\
KM 		&$0.938$&$0.465$& $1$     & $0.176$ &$0.228$ & $0.215$ \\
\hline
MR 		&$0.828$&$0.564$& $0.575$ & $1$     &$0.850$ & $0.824$ \\
DM 		&$0.803$&$0.497$& $0.613$ & $0.697$ &$1$     & $0.916$ \\
SA 		&$0.532$&$0.440$& $0.555$ & $0.649$ &$0.882$ & $1$ \\
\hline
\end{tabular}
\label{table2}
\end{table}

\begin{table}[t]\center
\caption{Same as table 2 for gene expression data. Note that
generally, if $\mathcal{S}$ has more clusters than $\mathcal{S'}$, one would
expect $O(\mathcal{S}|\mathcal{S'})<O(\mathcal{S'}|\mathcal{S})$, provided that
the distribution of cluster size is approximately the same. In this
example instead, $O(AL|\mathcal{S})>O(\mathcal{S}|AL)$, for $\mathcal{S}=$ KM,
MR, DM and SA, even if AL has more clusters than $\mathcal{S}$. This is due to
the fact that, at odds with configurations $\mathcal{S}$, AL has a very uneven
cluster size distribution with two large clusters ($n_s\approx
40$). The same happens for SL.}
\begin{tabular}{l|ccc|ccc|}
$\mathcal{S'}\backslash \mathcal{S'}$ & SL    &  AL   &  KM   &  MR   &  DM   &  SA\\
\hline
SL              & $1$   &$0.311$&$0.224$&$0.200$&$0.179$&$0.181$ \\
AL              &$0.204$& $1$   &$0.485$&$0.561$&$0.457$&$0.472$ \\
KM              &$0.118$&$0.390$& $1$   &$0.433$&$0.454$&$0.463$ \\
\hline
MR              &$0.075$&$0.322$&$0.310$& $1$   &$0.531$&$0.445$ \\
DM              &$0.086$&$0.335$&$0.416$&$0.678$& $1$   &$0.558$ \\
SA              &$0.076$&$0.302$&$0.369$&$0.495$&$0.487$& $1$   \\
\hline
\end{tabular}
\label{table3}
\end{table}

The geometric overlap is dominated by clusters which contribute
negligibly to the likelihood and which are therefore not statistically
significant. It is then preferable to introduce a likelihood overlap
\[
O_l(\mathcal{S'}|\mathcal{S})=\frac{\mathcal{L}_c(\mathcal{S}\cap\mathcal{S'})}
{\mathcal{L}_c(\mathcal{S})}
\]
which yields the fraction of likelihood of configuration $\mathcal{S}$
which is also ``explained'' by configuration $\mathcal{S'}$. Tables
\ref{table4} and \ref{table5} show that most of the relevant
information (in terms of likelihood) gained by standard methods is
also contained in ML configurations found with our method. The
opposite is however not true.

\begin{table}[t]\center
\caption{Likelihood overlaps $O_l(\mathcal{S'}|\mathcal{S})$ for financial
data. Values $O_l(\mathcal{S'}|\mathcal{S})>1$ imply that the intersection
of the two configuration is more likely than the configuration $\mathcal{
S}$.}
\begin{tabular}{l|cccccc}
$\mathcal{S}\backslash \mathcal{S'}$ & SL  &  AL   &  KM   &  MR   &  DM   &  SA\\
\hline
  SL & $1.000$  & $1.018$  & $1.006$  & $0.996$  & $0.990$  & $0.982$ \\
  AL & $0.643$  & $1.000$  & $1.057$  & $1.076$  & $1.084$  & $1.096$ \\
  KM & $0.509$  & $0.846$  & $1.000$  & $0.961$  & $0.967$  & $0.972$ \\
  MR & $0.485$  & $0.828$  & $0.924$  & $1.000$  & $0.974$  & $0.975$ \\
  DM & $0.474$  & $0.821$  & $0.915$  & $0.958$  & $1.000$  & $0.984$ \\
  SA & $0.468$  & $0.826$  & $0.915$  & $0.955$  & $0.980$  & $1.000$ \\
\end{tabular}
\label{table4}
\end{table}

The discrepancy between results of deterministic minimization and
hierarchical clustering algorithms is much reduced if one uses $\mathcal{
L}_c$. Indeed, MR and DM predict a number of clusters $K$ of the same
order whereas AL predicts a much larger $K$ than KM, specially if
correlations are weak, as in the financial data set (see table
\ref{table1}). As a further evidence, Fig. \ref{figovlp} compares the
overlap between configurations AL and KM with $K$ clusters with the
overlap between the configuration MR with $K$ clusters and the
corresponding configuration DM (see caption): Configurations MR and DM
are more similar than configurations AL and KM.

At odds with SL, AL and CL algorithms, hierarchical clustering yields
a ML configurations (MR) which is quite consistent with maximization
algorithms (SA or DM) which are computationally more demanding. This,
we believe, is due to the fact that $\mathcal{L}_c$ generates in typical
cases a configuration landscape which is smoother and with more easily
accessible maxima, than that generated by conventional distance-based
cost functions, such as $H_{KM}(\mathcal{S})$. This conjecture is
partially confirmed by looking at simple cases. Take, for example, a
data set generated by Eq. \req{ansatz} with constant correlation $\vec
x_i\vec x_j=g>0$ for all $i\neq j$. All configurations with two
clusters have the same cost $H_{KM}$, and there are $2^{N-1}-1$ such
configurations. But the configurations of maximal $\mathcal{L}_c$ with
two clusters are only those with an isolated object and a large
cluster of size $N-1$ (there are $N$ such configurations). Thus $\mathcal{
L}_c$ gives a much smaller degeneracy than $H_{KM}$ in this case.

\begin{figure}[h]
\centerline{\includegraphics[width=7cm,height=10cm,angle=270,clip]{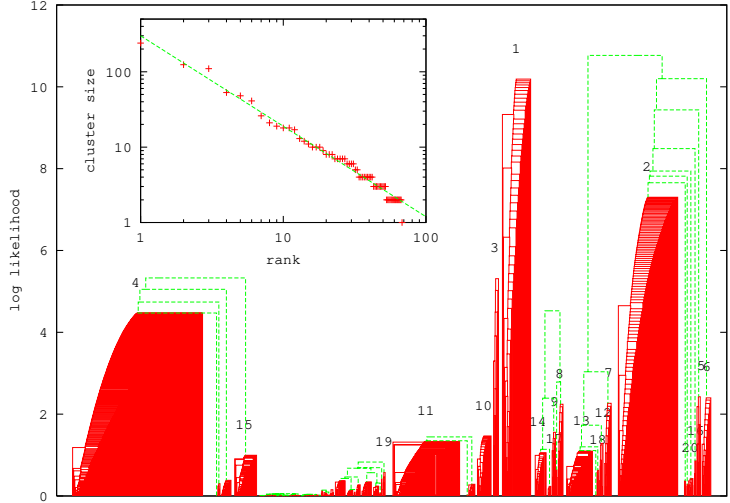}}
\caption{Dendrogram for financial data. The labels report the cluster
numbers: For example, cluster 1 contains firms in the electric sector,
cluster 3 is the sector of gold, 5 contains telecommunication firms, 6
is composed of banks, 8 and 14 petroleum, 9 oil and gas, 10
computers. Cluster 15 also lists firms in the energy sector. Clusters
2 and 4 are large mixed clusters. Inset: Rank plot of cluster sizes
for $\mathcal{S}_{MR}$. The dashed line (drawn as a guide to the eyes)
has slope $-1.2$ which implies that the number of clusters with size
larger than $n$ is proportional to $n^{-0.83}$. {The same plot for an
data set of $N=1000$ uncorrelated time series yields structures with a
log-likelihood of $\le 0.05$. The structure revealed in this plot is
on a scale of log-likelihood two orders of magnitude larger.}}
\label{figdendfin}
\end{figure}

\begin{table}[t]\center
\caption{Same as table 4 for gene expression data.}
\begin{tabular}{l|cccccc}
$\mathcal{S}\backslash \mathcal{S'}$ & SL  &  AL   &  KM   &  MR   &  DM   &  SA\\
\hline
  SL & $1.000$  & $1.192$  & $1.078$  & $1.201$  & $1.158$  & $1.149$ \\
  AL & $0.432$  & $1.000$  & $0.808$  & $0.925$  & $0.862$  & $0.843$ \\
  KM & $0.364$  & $0.753$  & $1.000$  & $0.736$  & $0.782$  & $0.761$ \\
  MR & $0.408$  & $0.866$  & $0.739$  & $1.000$  & $0.851$  & $0.792$ \\
  DM & $0.383$  & $0.788$  & $0.766$  & $0.830$  & $1.000$  & $0.810$ \\
  SA & $0.376$  & $0.761$  & $0.737$  & $0.763$  & $0.801$  & $1.000$ \\
\end{tabular}
\label{table5}
\end{table}

\begin{figure}[t]
\centerline{\includegraphics[width=7cm,height=10cm,angle=270,clip]{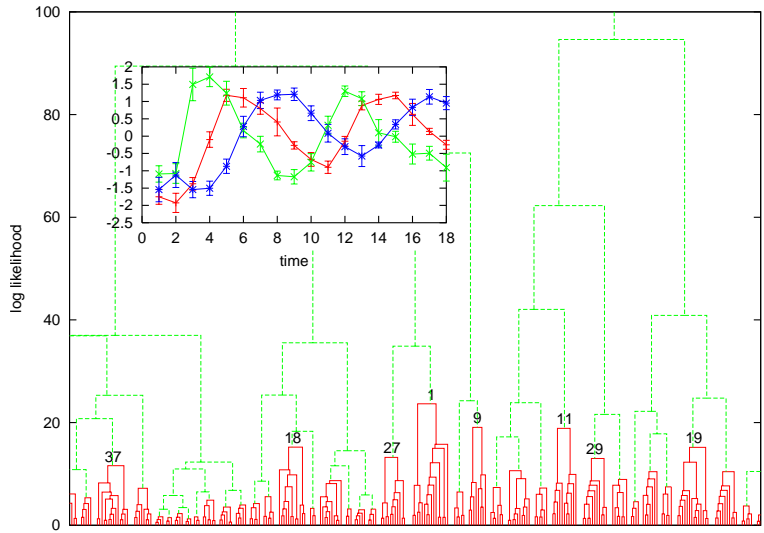}}
\caption{A section of the dendrogram for gene expression data. The
labels report the cluster numbers; smaller number indicate higher
likelihood. Inset: average patterns of (log)expression in the first
three clusters. Error bars represent the variance inside a
cluster. Observe how the clusters identify different patterns with
small dispersion. {Note that the scale of log-likelihood is an order of
magnitude larger than that in Fig. \ref{figdendfin} (i.e. 3 orders
of magnitude larger than the noise background).}}
\label{figdendgene}
\end{figure}

The MR algorithm delivers information on the hierarchical organization
of the data set. The dendrograms for the two case studies are shown in
Figs. \ref{figdendfin} and \ref{figdendgene}. The dendrogram of market
data reveals a complex structure consistent with the scaling laws
reported in Ref. \cite{GM} (see inset). A much more uniform cluster
distribution is found in gene expression data. Figs. \ref{figdendfin}
and \ref{figdendgene} only report the first two types of branch
points: {\em i)} those which increase the total likelihood and {\em
ii)} those which do not increase the total likelihood but where the
likelihood of the unified cluster is larger than those of both
individual clusters. These two sets of branching points are clearly
separate and describe the cluster structure at different
resolutions. In most branching of type {\em i)} a large cluster merges
with a single object whereas type {\em ii)} steps merge clusters into
larger clusters.

\begin{figure}[t]
\subfigure
{\includegraphics[width=8cm,height=6cm,clip]{fig5a.eps}}
\subfigure
{\includegraphics[width=8cm,height=6cm,clip]{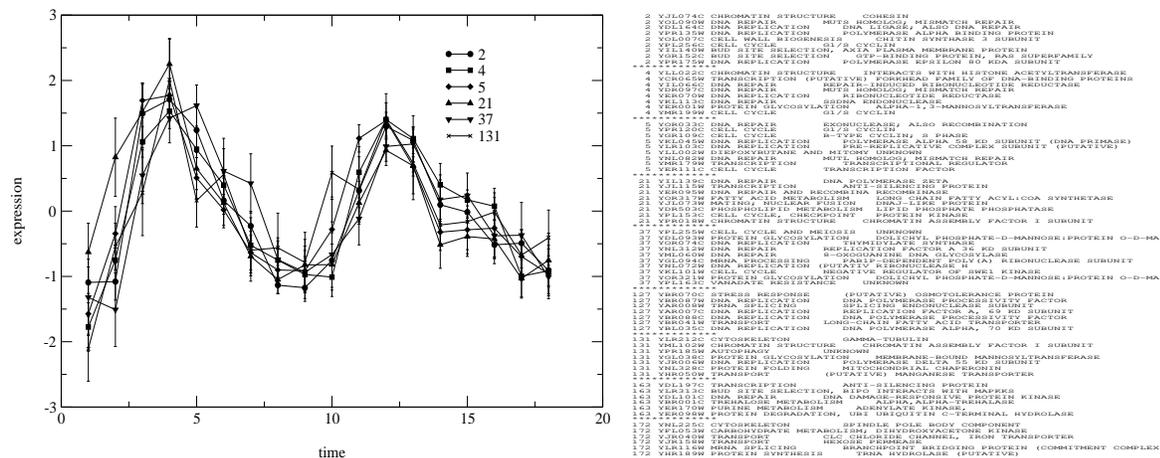}}
\caption{The first 6 elements of cluster number $2$ obtained from the
second step of reclustering, and the composition of the same
cluster. In the table, the first column indicates the number of the
cluster after the first step, the second the code of the gene, and the
third the corresponding function. At this level of description almost
all the information available has been exploited. Observe for example
how genes associated to DNA repair, which are put in different
clusters after the first step, are found now in the same cluster.}
\label{clust2}
\end{figure}

Gene expression data shows a rich hierarchical structure of type {\em
ii)} branchings. The ML structure has a large number of clusters with
$8\div 15$ elements. This suggests that the model Eq. \req{ansatz}
does not provide an exhaustive description of the correlations of this
data set. There are indeed sizeable correlations between clusters that
are not taken into account. An alternative way to describe these
correlations, besides that provided by the MR algorithm, is to perform
a {\em reclustering} of the cluster patterns $\vec X_s$ with the SA
algorithm. More precisely this amounts to taking $\vec x_s=\vec
X_s/\|\vec X_s\|$ as the input data set. Reclustering steps can be
iterated until a configuration with singleton clusters is found. This
leaves us with a sequence of nested reclustered configurations which
provide a hierarchical description of the correlations present in the
original data set. In fig. \ref{clust2} we show the effect of a second
application of the SA clustering procedure on the biological data set.

\section{Conclusions} 

We have discussed a maximum likelihood approach to data clustering and
its application to two examples. The method is based on a simple
statistical description of data where similar objects have something
in common (Eq. \ref{ansatz}). The likelihood that a particular data
set is described by such a model with a given cluster structure can be
efficiently computed. This provides a parameter-free measure for
cluster structures which can then be taken as the basis of clustering 
algorithms. 

Different algorithms to find maxima of the likelihood -- or good
approximations to them -- have been introduced. On one hand we discuss
computationally expensive methods, such as simulated annealing
technique (SA) and deterministic minimization (DM) algorithms, which
are expected to provide a good approximation to the maximum likelihood
structure. At odd with most other methods, the number of clusters
should not be fixed {\em a priori} but is predicted by our algorithms.
On the other hand, we introduce a simple deterministic hierarchical
clustering algorithm (MR) which provides a much faster approximation.

We show that standard methods with distance based measures of
clustering predict cluster structures which may be very different
according to the specific algorithm used (SL, AL or KM). On the
contrary, the maximum likelihood structures predicted by different
algorithms based on the measure $\mathcal{L}_c$ (MR, DM and SA) are very
similar. This suggests that, for instances where a meaningful cluster
structure exists, the {\em log-likelihood landscape} has a broad and
easily accessible maximum, whereas distance-based cost functions such
as $H_{KM}$ (Eq. \ref{HKM}) may give rise to a more complex landscape.

Our approach is particularly suited for large dimensional data
sets. Indeed, the dimensionality $D$ only enters in the calculation of
Pearson's coefficients $C_{i,j}$.

Data clustering has been regarded as an ill defined problem (see
e.g. \cite{ill}). Indeed in conventional approaches one needs to
define a similarity measure and a resolution scale. The approach
proposed in the present work does not suffer from these drawbacks. The
cluster structure is entirely determined by the internal correlations
of the data set. This makes data clustering a well defined problem.

We are grateful to R. N. Mantegna for providing the financial data
set.


\end{document}